\documentclass[12pt]{article}
\usepackage{makeidx}
\usepackage{graphicx}
\usepackage{subfigure}
\usepackage{multicol}
\usepackage{amssymb}
\usepackage{amsmath}
\usepackage{tabularx}
\usepackage{multirow}
\usepackage{array}
\usepackage{color}

\newenvironment{sciabstract}{%
\begin{quote} \bf}
{\end{quote}}

\newcounter{lastnote}

\title{Nonlinear Decelerator for Payloads in Aerial Delivery Systems. I: Design and Testing$^{}$\footnote{Preprint available at \textit{arxiv.org/submit/1045587/2014}, submitted to \textit{Nonlin. Dyn.}}}
\author
{T. Lyons,$^{1}$ M. Ginther,$^{2}$  P. Mascarenas,$^{2}$ E. Rickard,$^{2}$ \\
J. Robinson,$^{2}$ J. Braeger,$^{3}$ H. Liu,$^{1}$ and A. Ludu$^{1}\footnote{Correspondence should be addressed to ludua@erau.edu.}$\\
\\
\normalsize{$^{1}$Department of Mathematics, Embry-Riddle Aeronautical Univeristy,}\\
\normalsize{Daytona Beach, FL 32114, USA}\\
\normalsize{$^{2}$Adams State University, Alamosa, CO 81101, USA} \\
\normalsize{$^{3}$Embry-Riddle Aeronautical University,}
\normalsize{Prescott, AZ 86301, USA}\\
\\
}
\date{}
\begin{document}
\baselineskip12pt
\maketitle
\begin{sciabstract}
\centerline{Abstract}
\vskip 1cm
We study the dynamics and the optimization of the shock deceleration supported by a payload when its airborne carrier impacts the ground. We build a nonlinear elastic model for a container prototype and an elastic suspension system for the payload. We model the dynamics of this system and extract information on maximum deceleration, energy transfer between the container and payload, and energy resonant damping. We designed the system and perform lab experiments for various terminal velocities and types of grounds (cement, grass, sand water, etc.). The results are compared with the theoretical model and results are commented, including predictions for deceleration at different types of ground impact. The results can be used for aerial delivery systems, splash-down of capsules, recoveries,  weather balloons, coastal surveying systems, or the new introduced goal-line technology in sport competitions.
\end{sciabstract}

\vfill
\eject

\section{Introduction}
\label{sec:sec1Intro}

Aerial delivery systems, \cite{snowflake}, can be described as airborne compact containers, sometimes attached to a parachute, containing a payload consisting in analog and digital recording equipment for collecting data at high altitude. Such systems can stream the recorded data into a satellite network, or just store it and aim for further recovery on the ground. Given the fact that such system are in general expensive, and the data stream consists in a very large volume sent at very high frequency (like multi-spectral high definition video recordings), recovering these systems is a desired procedure.

In general, after accomplishing the data recordings, the air delivery systems are disconnected from their aerial carrier, and released airborne with a parachute. Even in this situation the ground impact can be seriously damaging for the sensitive payload, especially since it is difficult to determine the landing on a specific predestinated "soft" spot. The study of the maximal decelerations induced in the payload at their ground impact represents an interesting field of study. Moreover, a large spectrum of ingenious mechanical systems can be used to suppress these decelerations to an acceptable limit, no matter of the type of landing, be it on concrete, grass or water, etc.

The system container plus payload can be design to minimize this ground impact deceleration by using combinations of nonlinear normal modes from special elastic systems. One interesting solution is to use the non-linear energy sink process (also called targeted energy transfer) in order to transfer quickly, by transient resonance capture, the shock energy from the payload back to the container oscillations \cite{nes1,bookNMLNS}.

Such a design capable of quickly damping the payload deceleration can be further used in a variety of different research projects including weather balloons, coastal surveying systems, goal-line technology in sport competitions, splash-down of capsules, etc.

In this paper we present an air delivery system prototype capable of reduced payload deceleration touchdown for a variety of ground types and terminal velocities. Several such systems have been designed and tested thus far. The goal of this research is to study what is the best protection system of the payload inside the container against the mechanical damaging effects of the high value of deceleration resulting from the ground impact.

Section \ref{sec:sect2generalimpact} begins by providing a brief description of the system and measurement methods. In continues by describing the design of the container and the payload, and the mathematical model. The container is analyzed in parallel by two methods: thin shell model modes of oscillations, and Hertz elastic deformation. The first model generates the normal frequencies of vibrations, and the second model generates a nonlinear force of deformation, including a phase-transition. Exact solutions of the model equation for oscillations under this nonlinear deformation force are compared with the thin shell modes in order to obtain a reliable combined model for the elastic behavior of the container. The payload is also modeled by a system of springs whose force combines into a nonlinear interaction. The model equation for this force is also integrable, but  the oscillation obtained are too tedious to be studied exactly. We use a series expansion in order to calculate with a reasonable approximation the frequency of oscillation of this nonlinear force.

In section \ref{sec:impactmodeling} we present the theoretical model and its performances. We also compare the theoretical results with the experiment and calculate rates of dissipation of energy.

In section \ref{sec:last} we analyze the results of the experiments  on drops performed at different terminal velocities on different types of ground. The paper ends with conclusions and recommendations for the further development.

\section{Container and payload}
\label{sec:sect2generalimpact}

In order to understand the dynamics of the impact between the compound system container plus payload with the ground, and to predict maximal decelerations for different terminal velocities and different types of ground we performed controlled experiments of free drops in the Wave Motion Lab at ERAU \cite{report}. For the typical range of container sizes used in aerial data acquisition, and  corresponding parachutes, the terminal velocities in normal conditions ranges in $V_0 \sim 3-8$ m/s. This velocity can be reached in free fall without parachute inside the controlled atmosphere of a lab from heights not exceeding $5$ m. In our experiments the container was dropped from different heights ranging from 1 cm to 4 m on different types of landing floors: cement, grass, sand, and water.

\begin{figure}[!b]
\centering
\centering\includegraphics[scale=0.1]{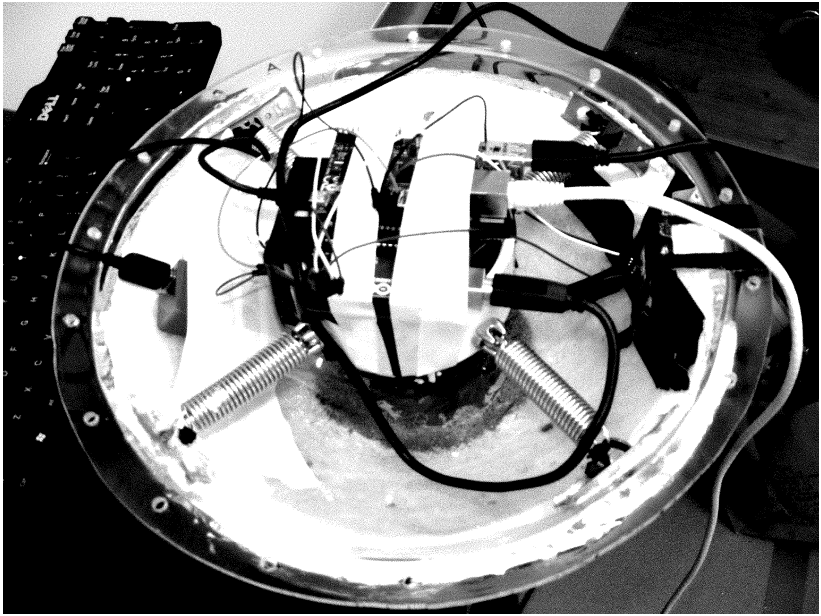}
\caption{View of the opened container during reading data after one drop. Four radial springs connect the container's wall to the cex
ntral payload which contains the video camera, accelerometers and the integrated circuit board. One accelerometer is placed on the container's wall, too.}
\label{fig:system}
\end{figure}

The delivery system, Fig. \ref{fig:system} shows the payload container, chosen
to be a watertight, crushproof, and dust proof acrylic custom made oblate spheroid case with the measurement and video recording units occupying a small fraction of its interior chamber.

\begin{figure}
\centering
\begin{tabular}{cccc}
{\includegraphics[width=7cm,height=7cm]{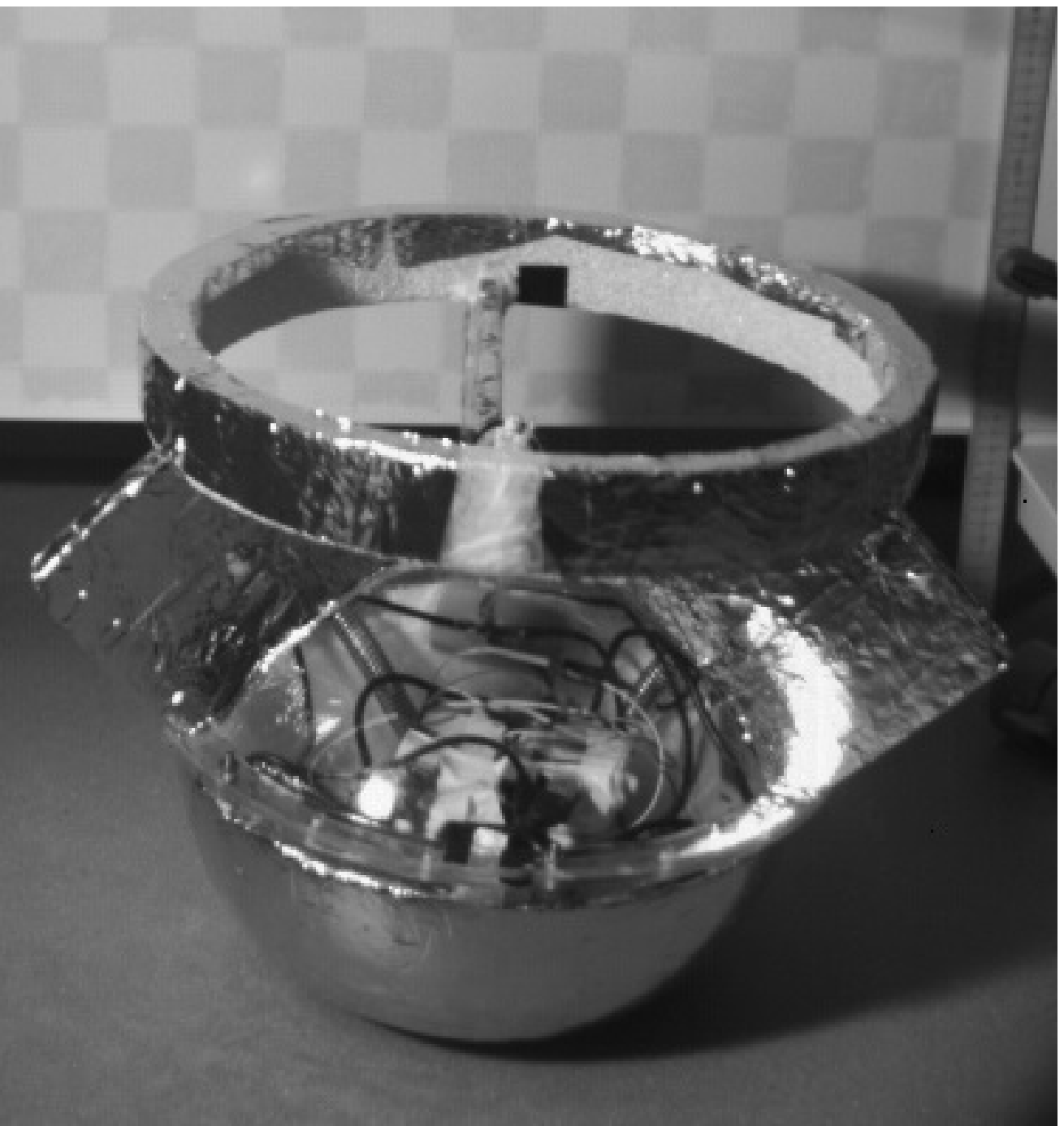}} &
{\includegraphics[width=7cm,height=7cm]{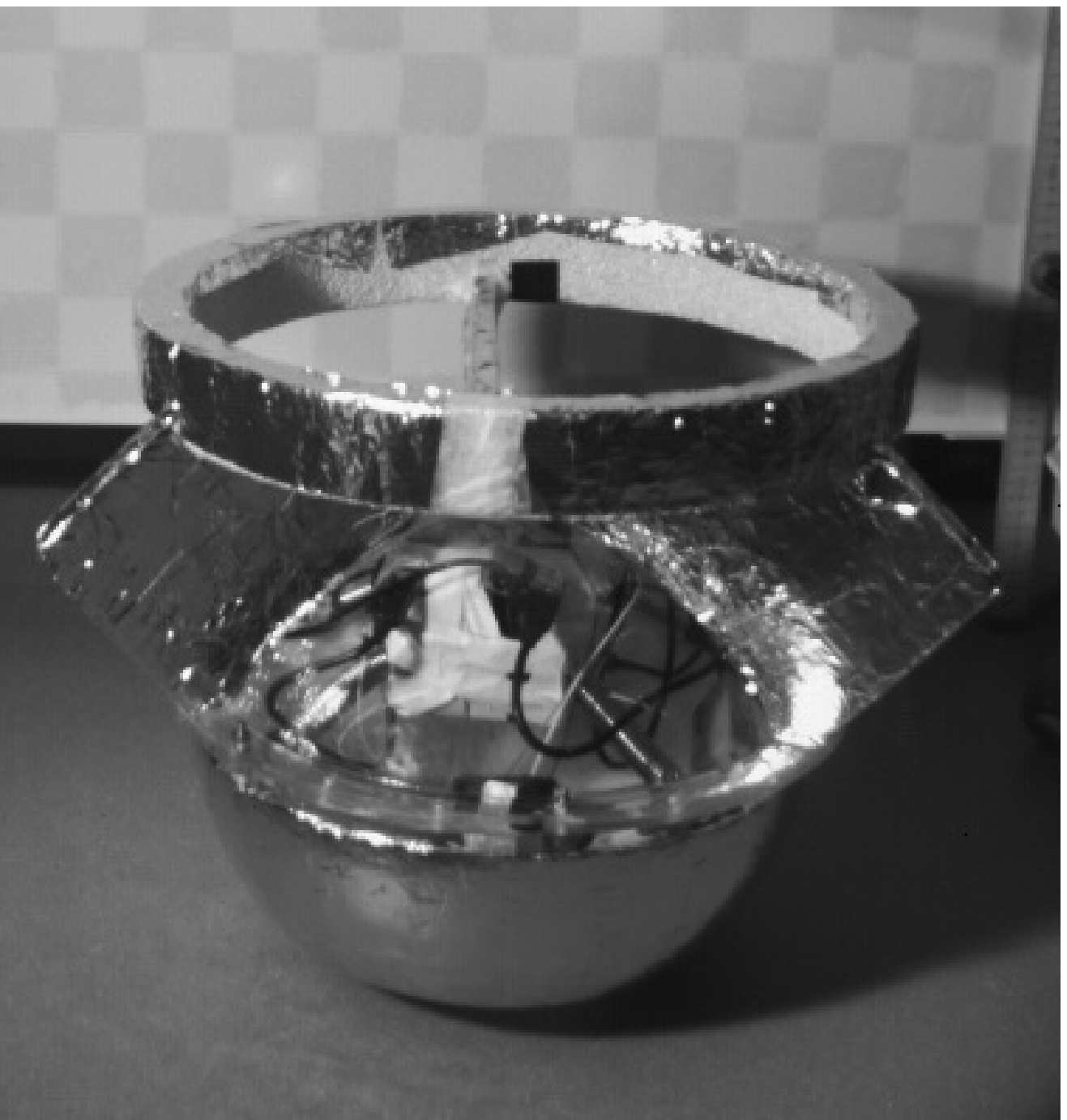}} \\
{\includegraphics[width=7cm,height=7cm]{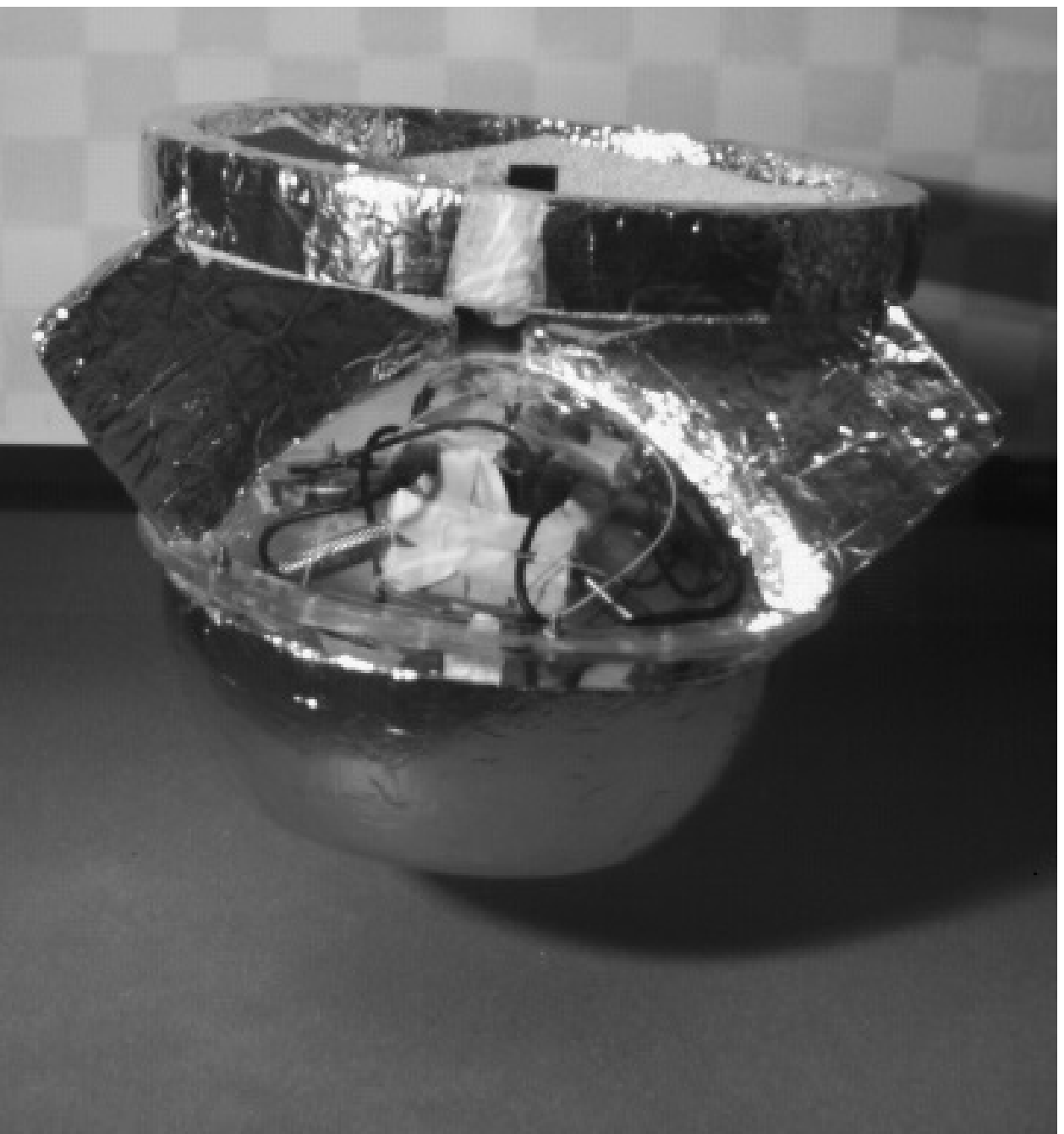}} &
{\includegraphics[width=7cm,height=7cm]{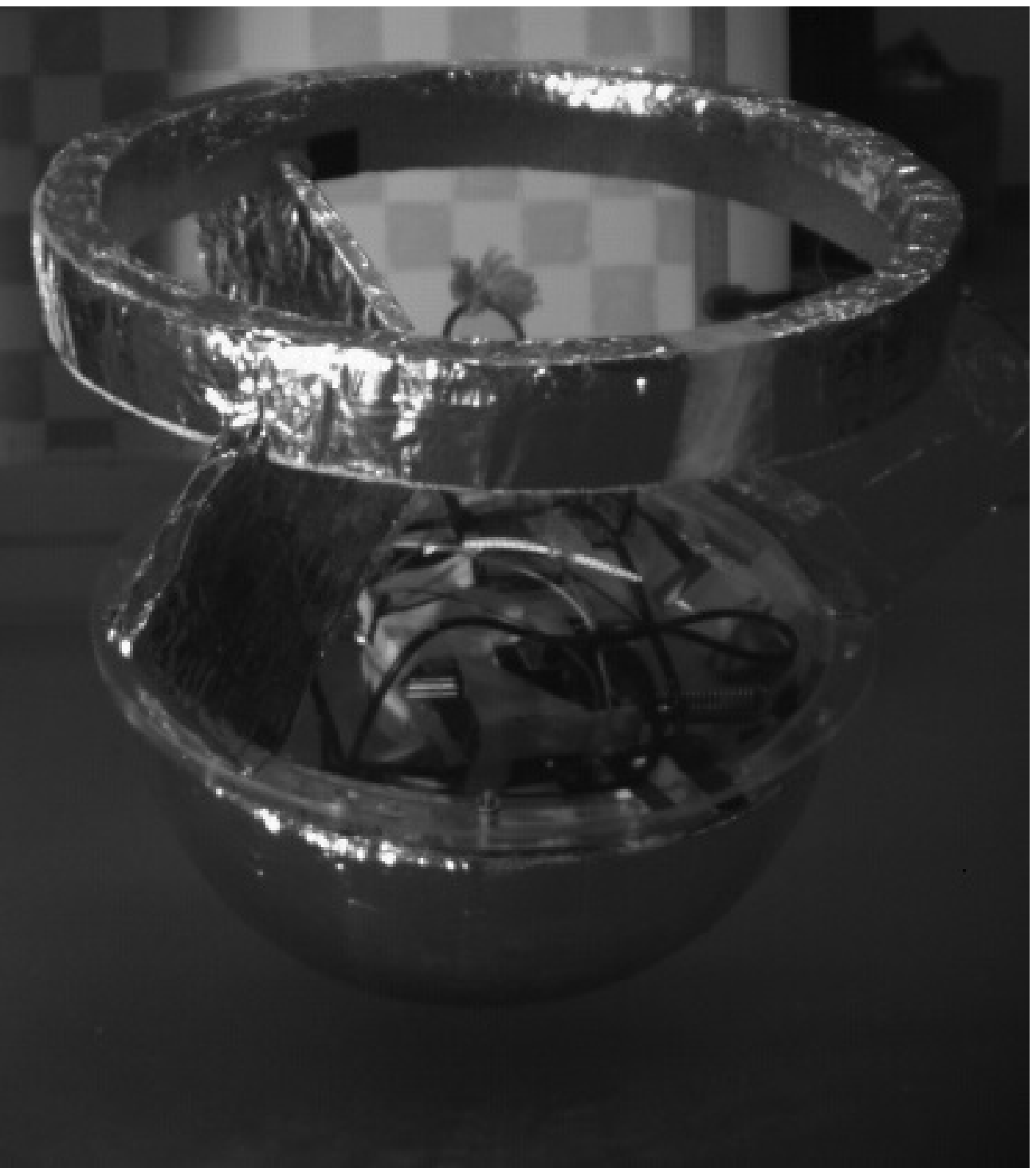}} \\
\end{tabular}
\caption{Impact after $H=2.0$ m drop on concrete. From top left clock-wise: (a) First impact: the payload (white square box inside) sinks downwards; (b) While container is still on ground the payload bounces up and extends the springs upwards to its maximum relative to container height; (c) The container is pulled upwards by the moving payload; (d) The payload swings downwards in opposition to the lifting motion of the container.}
\label{fig:rapidphotocexp3concr}
\end{figure}

The accelerations of the container and the payload inside were measured with a combination of a MPU-6050 3-axis gyroscope and 3-axis accelerometer type 6-DOF Module MMA7361, and an accelerometer sensor module Speed A891NN. The position versus time of the container was recorded with an H-EM 501 high speed streaming AOS camera with Cannon lenses. Structurally, the container built by \textit{Hydroplus Engineering} has an oblate spheroidal shape having its polar axis (along the vertical direction) $2R_{\|}=0.24 m$ and its horizontal diameter $2 R_{\bot}=0.32m$. The container is made of two symmetric halves of transparent Acrylite GP material of thickness $h=4.78 $ mm interconnected by a bolted rim through a water proof gasket of width $0.02m$ outstanding the spheroid surface. On top of its upper half the container has an drag-producing empennage made of four vertical fins connected by a horizontal circular stabilizer, see Figs. \ref{fig:system}, \ref{fig:rapidphotocexp3concr}.

\subsection{Container model}
\label{subsection:container}

In order to model the container impact we need to study first the elastic properties of the container. One way to perform this analysis is to use the modal response of a thin spherical shell \cite{thinshellVibr1,thinshellVibr2,viscoelastic}. The axisymmetric vibration modes spectrum is basically composed of three countable sets of frequencies: one for the membrane, one for the flexural, and one higher  mode. The non-axisymmetric modes have the same frequencies except in a degenerate way. The modal frequencies  for the thin spherical shell are given by \cite{thinshellVibr1}
\begin{equation}\label{eq:freqThinShellModes}
f_{i}=\frac{\lambda_{i}}{2\pi R}\sqrt{\frac{E}{\rho (1-\nu^2)}}, \ i=1,2, \dots,
\end{equation}
where $R$ is the sphere radius, $E$ is the modulus of elasticity, $\nu$ is the Poisson ration and $\rho$ is the density. The parameters $\lambda_{i}$ are given by the roots of the shell model equation \cite{thinshellVibr3}:
\begin{equation}\label{eq:frequencies}
\alpha_{i} \lambda^6-\beta_{i} \lambda^4+\delta_{i} \lambda^2-\gamma_{i}=0.
\end{equation}
The coefficients for this equation depend on a mode number $i$, on the size $h,R$ through the coefficient $h^2 / 12 R^2$, and on $\nu$. In the case of our material we have $\rho=1.19 \cdot 10^3$ $Kg/m^3$, $E=4.75 \cdot 10^7$ $N/m^2$,  $\nu=0.35$, $R=R_{\|}$ and $h$ as above. The frequencies of the first modes are shown in Table \ref{tab:eee}, where the three rows represent the three double real roots of Eq. (\ref{eq:frequencies})
\begin {table}[H]
\caption {Theoretical normal modes of oscillations (frequencies) of the container sphere used in experiments and simulations.} \label{tab:eee}
\begin{center}
\begin{tabular}{|c|c|c|c|c|c|}
\hline
$i$ & $1$ & $2$ & $3$ & $4$ & $5$ \\
\hline
$f(i) [Hz]=$ & - & 22.1 & 26.3 & 28.1 & 29.2 \\
\hline
$f(i) [Hz]=$ & 65.3 & 89.4 & 118 & 150 & 181 \\
\hline
$f(i) [kHz]=$ & 2.93 & 2.94 & 2.94 & 2.95 & 2.96 \\
\hline
\end{tabular}
\end{center}
\end{table}
The main modes of vibrations of the container have periods in the range $35-45$ ms for the flexural modes (the first row in the table), and $5.5-15$ ms for the membrane modes (the second row), while the highest modes (third row) cannot be excited by the impact initial conditions and at energies in the range of the impact with ground. Studies of correlations of theoretical natural frequencies of spherical shells with finite element simulations and with time-averaged dynamic holography  experiments with real spheres with imperfections show,  \cite{thinshellVibr3}, that the non-axisymmetric modes splitting towards higher values of frequency, and the most likely modes of excitation are the short period membrane modes.

Another way of studying the elastic deformation of the bottom of the container is to use the Hertz elastic hypothesis, \cite{elastica},  where energy of deformation given by
\begin{equation}\label{eq:Hertz}
U=E \int \biggl( \frac{\partial u}{\partial z} \biggr)^2 dV,
\end{equation}
where $u$ is the strain field and $dV$ is the volume element. This approach seems to be appropriate for deformations induced only by bottom impacts where the local strains  are rather one-dimensional induced by the vertical motion of the spheroidal container. This hypothesis is also supported by a multitude of rapid photography data and measurements of rotation and lateral acceleration. Indeed, since the bottom of the container, and the payload suspension are built symmetrically off-axis lateral motion and rotations during the free fall and impact phases with a plane surface phases are very unlikely.

\begin{figure}[!b]
\centering
\includegraphics[scale=0.6]{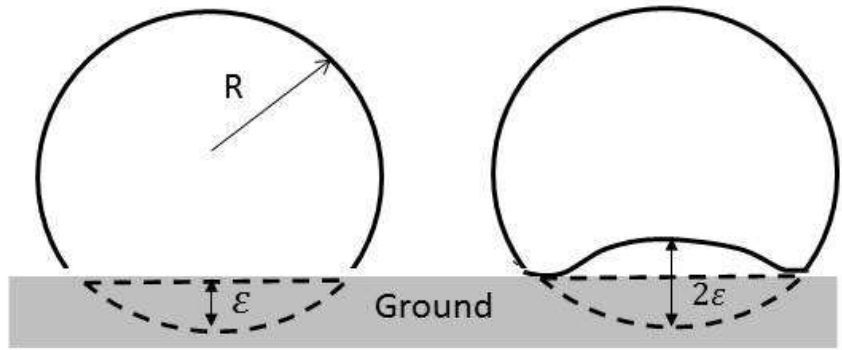}
\caption{Configurations of small axial deformation of a thin spherical shell of radius $R$ at impact on a rigid horizontal plane. \textit{Left:} (I) Flat deformation of height $\epsilon$. \textit{Right:} (II) Fold and trough deformation occurring usually at double amounts of deformation than the flat one. At higher impact force the bottom of the sphere buckles upwards and builds an inversion of curvature.}
\label{fig:impactball}
\end{figure}

It was shown in \cite{pingpongBall} that the impact of an elastic thin sphere with a rigid flat surface can be described as a two-phases process. In the first phase (I) the shell flattens against the horizontal surface, the impact is weakly nonlinear and the applied force is relatively small. The second phase (II) occurs if the compression continues above this limit with a higher force. In this phase the flattened region  buckles upwards, and this inversion of curvature leads to a circular fold and a trough, see Fig. \ref{fig:impactball}.

Experiments on elastic spherical shells, \cite{pingpongBall}, show that the flattened shape springs to the buckled configuration suddenly revealing a first-order phase transition at a deformation close to twice the thickness $h$ of the shell in average.
In any of the phases shown in Fig. \ref{fig:impactball} the elastic deformation energy has two terms. One term represents the compression of the bottom spherical cap into a flat area (type I), or into a concave trough shape (type II). The other term describes the bending, i.e. the occurrence of a circular fold around the compressed domain. If we denote the axial deformation along the symmetry axis by $\epsilon << R$, the compression part of the elastic deformation energy Eq. (\ref{eq:Hertz}) can be evaluate by simple trigonometric calculations \cite{pingpongBall}
\begin{equation}\label{eq:firstpartelastenrg}
U_{comp}=\frac{\pi E h \epsilon^3}{144 R}.
\end{equation}
For the bending part of the elastic energy we use the theory of equilibrium of thin plates, \cite{elastica}, and following the Willmore energy formula, \cite{elast2}, we obtain
\begin{equation}\label{eq:thinplatesenergyelast}
U_{bend}=E h \int \biggl( \frac{h^2 H^2}{24 (1-\nu^2)}+\bar{\kappa} K \biggr) dA,
\end{equation}
where $H, K$ are the mean and Gaussian curvatures, $dA$ is the area element, and $\bar{\kappa}$ is the Gaussian curvature modulus \cite{elast3}. In the limit of thin shell model, from Eqs. (\ref{eq:firstpartelastenrg},\ref{eq:thinplatesenergyelast}) we obtain the elastic energy $U=U_{comp}+U_{bend}$ in the form \cite{pingpongBall}
\begin{equation}\label{eq:energymorefinal}
U(\epsilon)=\left\{
\begin{array}{rl}
U_{I}= \frac{E \pi h^{5/2}\epsilon^{3/2}}{30 [60 (1-\nu^2)^5]^{1/4}R}+\frac{\pi E h \epsilon^3}{144 R} & 0< \epsilon \leq \epsilon_{c}, \\
U_{II} = \frac{2E \pi h^{5/2}\epsilon^{3/2}}{15 [60 (1-\nu^2)^5]^{1/4}R}+\frac{\pi E h^3 \epsilon}{144 R} & \epsilon_{c} < \epsilon,
\end{array} \right.
\end{equation}
where $\epsilon_{c}$ is the smallest positive solution of the equation $U_{I}(\epsilon)=U_{II}(\epsilon)$, and it represents the point of phase-transition from the flat deformation to the trough deformation. The second row in Eq. (\ref{eq:energymorefinal}) differs from the first by a factor of 4 in the first term, and the linear dependence on $\epsilon$ in the last term.

For small deformations the expression $I$ is energetically favorable. However, as the deformation increases $U_{II} < U_{I}$  and a first order phase-transition arises at this critical deformation roughly proportional to $h$. The hysteresis associated to this transition (the force has a jump) is strongly dissipative because of the friction work generated by the sliding  of the contact point between the circular fold and the ground, when the radius of the fold changes. For a friction coefficient $\mu$ between the container material and ground, the dissipated energy can be approximated with
\begin{equation}\label{eq:dissipatedenergy}
W= \frac{E \mu \pi h^{5/2}\epsilon}{20 [15 R^2 (1-\nu^2)^5]^{1/4}} ,
\end{equation}
showing is a linear dependence with the deformation.

Another interesting feature of the Hertz model is that we can estimate the impact time
from the impact velocity with pretty good accuracy
\begin{equation}\label{eq:impacttime}
\tau=\biggl( \frac{|V_{0}|}{c} \biggr)^{1/3} \frac{R^2}{hc},
\end{equation}
where $c$ is the speed of sound in the wall material and $V_{0}$ is the initial impact velocity. In our case we measured $c=2,745$ m/s, and for example for an impact at $V_{0}=-2$ m/s it results an impact time $\tau\sim 62$ ms, which is in  good agreement with our measurements of the impact, and with the dynamical estimation, see the following sections.

The elastic impact forces obtained from Eqs. (\ref{eq:energymorefinal}) have the form
\begin{equation}\label{eq:forcesimpact1}
F(\epsilon)=\left\{
\begin{array}{rl}
-a \sqrt{\epsilon}  -b \epsilon^2, \ \  0 \leq \epsilon \leq \epsilon_{c},  \\
-4 a \sqrt{\epsilon}  -c, \ \  \epsilon_{c} < \epsilon,
\end{array} \right.
\end{equation}
with the positive constants $a,b,c$ obtained from Eqs. (\ref{eq:energymorefinal}).

In the absence of damping the dynamics induced by each of these two types of forces is exact integrable through $_{2}F_{1}$ hypergeometric functions. For the type $I$ force, the resulting nonlinear oscillator model $m Z''=-a \sqrt{Z}-b Z^2$, with initial conditions $Z(0)=0,Z'(0)=-V_{0}$ has an implicit exact solution in the form
$$
t^2=\frac{3 m Z^2 (1-Q_{1} Z^{3/2})(1-Q_{2} Z^{3/2})F_{1}\biggl( \frac{2}{3}; \frac{1}{2}, \frac{1}{2}; \frac{5}{3}; Q_{1} Z^{3/2}, -Q_{2} Z^{3/2} \biggr)^2}{3 V_{0}^{2} m-4 a Z^{3/2}-2 b Z^3},
$$
where $F_{1}(a;b_{1},b_{2};c;x,y)$ is the Appel hypergeometric function of two variables, and
$$
Q_{1,2}=\frac{b}{-a\pm \sqrt{a^2+ 3 b m V_{0}^{2}/3}}.
$$
For the type $II$ force, the corresponding nonlinear oscillator model $m Z''=-4a \sqrt{Z}-c$, with initial conditions $Z(t_{c})=Z_{c},Z'(t_{c})=-V_{c}$ has an implicit exact solution in the form
$$
\frac{\Xi_{32} Z_{1} Z_{2} Z_{3} \biggl[ \Xi_{13} E\biggl( \sin^{-1} \sqrt{\frac{Z_{3}}{\Xi_{23}}}\biggl| \frac{\Xi_{23}}{\Xi_{13}}\biggr) -\Xi_1 F\biggl( \sin^{-1} \sqrt{\frac{Z_{3}}{\Xi_{23}}}\biggl| \frac{\Xi_{23}}{\Xi_{13}}\biggr) \biggr]^2}{\Xi_{13} \Xi_{23} \biggl( C_1 -2 c Z-\frac{4 Z^{3/2}}{3} \biggr) }=\pm\frac{t+C_2}{16 \sqrt{m}},
$$
where $F(\cdot | \cdot ), E(\cdot | \cdot )$ are the complete elliptic integrals of the first and second kind, respectively, and $C_{1,2}$ are constants of integrations. We define the symbols $\Xi_{ij}=\Xi_{i}-\Xi_{j}, Z_{i}=\sqrt{Z(t)}-\Xi_{i}$, where $\Xi_{i}, i=1,2,3$ are the solutions of the algebraic  equation $-3C_1+3c \Xi^2+8 a \Xi^3=0$.
\begin{figure}[!b]
\centering
\includegraphics[scale=1]{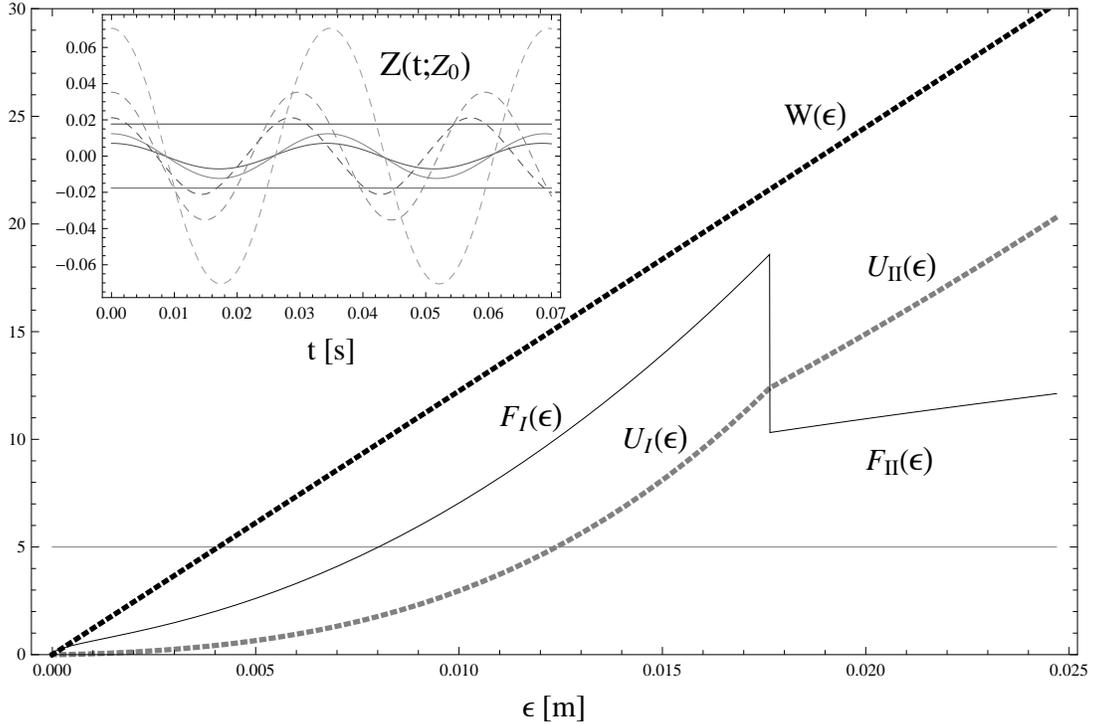}
\caption{The elastic energy $U$ (solid curve, Eq. (\ref{eq:energymorefinal})), the friction energy $W$ (dashed line, Eq. (\ref{eq:dissipatedenergy})), and the force $F$ scaled $1:100$ (dotted, Eq. (\ref{eq:forcesimpact1})). In the inset we present exact oscillation solutions for this force, for $V_0 =0$ and various initial positions $Z_0$: solid curves for $Z_0 < \epsilon_{c}$, and dashed curves for $Z_0 >\epsilon_{c}$. The two horizontal solid lines in the inset represent $Z=\pm \epsilon_{c}$. In spite of the discontinuity in force, the solution is smooth. Nonlinearity is noted in the dependence of the oscillation period with initial conditions. The dashed horizontal line is the maximum impact force measured by the acceleration of the container.}
\label{fig:pingpongmodel}
\end{figure}

In Fig. \ref{fig:pingpongmodel} we present a numeric example.  The elastic potential energy $U$ from Eq. (\ref{eq:energymorefinal}), the dissipated energy $W$ from Eq. (\ref{eq:dissipatedenergy}), and the resulting  force $F$ from Eq. (\ref{eq:forcesimpact1}) are plotted together with some exact oscillation solutions for various initial positions $Z_0, Z_{0}^{'}=0$, for a $m=4 Kg$ container made from the specified material. The jump in the force, and non-differentiability of energy are visible at the phase transition point which occurs at $\epsilon_{c}=0.0176$ m. Interesting enough, the jump--otherwise observed experimentally, yet not with such a strong discontinuity--in the force does not prevent solutions to be smooth. For $Z_0 < \epsilon_{c}$, the period of oscillations does not change too much with initial conditions. However, for $Z_0 >\epsilon_{c}$ the signature of the nonlinearity is noted through the strong dependence of the oscillation period with initial conditions.

From our fast photography experiments of dropping on concrete, and for heights $H=1 \div 3$m, we know that $Z$ or $\epsilon$ does not exceed $20$ mm which proves that the $\epsilon_{c}$ phase transition limit is not reached for this type of situations. Comparing the elastic energy and friction dissipation from Fig. \ref{fig:pingpongmodel} at this maximum values of deformation with the initial mechanical energy in the drop $39.24$ J at $H=1$m, and $58.86$ J at $H=1.5$m reveals the fact that only half or less of mechanical energy is lost friction with the ground, i.e. $35$\% energy loss for $H=1$ m drop, and $42$\% energy loss for drop at $H=1.5$ m. The rest of the energy is lost in visco-elastic deformation of the container material which will be consider in the following. The value of the impact force measured by the accelerometer attached to the container shows for $H=1.5$ m a maximum values on concrete of $465$ N. Form Fig. \ref{fig:pingpongmodel} it can be inferred that such a force involves deformation in the range $\epsilon=0.008 $m which is in full agreement with the theoretical oscillations shown in the inset of the same figure for $\epsilon < \epsilon_{c}\sim 0.0176$ m.

The period of oscillations in this sub-critical range of amplitudes is  $33.2$ ms which is exactly the $4^{th}-5^{th}$ mode predicted by the thin shell model, see the table above. Small differences from the thin shell model, and the tendency to excite higher shell mode may be related to the internal viscous and frictional forces in the container material which were not taken into account in the thin shell model. Also,  for deformations larger than $\epsilon \sim 0.005$ m the quadratic term in the Hertz force of type I in Eq. (\ref{eq:forcesimpact1}) ($F_{I}$) is dominant over the square root. In this case, the solution for the oscillations $m Z'' \sim -b Z^2$ is given by the Weierstrass elliptic function $\wp ((b Z/6)^{1/3};0,2T)$ which provides smaller values for the period of oscillations (stronger interaction).

\subsection{Payload}
\label{subsection:payload}

The payload consists in a square shaped solid box containing electronics, sensors, batteries and a video camera pointing downwards, see Fig. \ref{fig:system}. It is suspended at the center of the container by four linear springs of equilibrium length $l_{0}$ and elastic constant $k$ which allow it to oscillate up and down inside the container, see Fig. \ref{fig:rapidphotocexp3concr}.  If we denote by $z$ this vertical displacement relative to the container, the total vertical force has the expression
\begin{equation}\label{eq:forceModel4Springs}
f(z)=-4k z \frac{\sqrt{z^2+l_{1}^{2}}-l_0}{\sqrt{z^2+l_{1}^{2}}},
\end{equation}
where $l_{1}$ would be the extension of each spring when a massless payload stays in equilibrium in the equatorial plane. This nonlinear force has a linear term if the springs are pre-tension, that is if $l_{1} > l_{0}$.
The series expansion of the nonlinear force is
\begin{equation}\label{eq:seriesForceSprings}
f(z)= -4k \biggl( 1-\frac{l_{0}}{l_{1}}\biggr) z -2k \frac{l_{0}}{ l_{1}^{3}}z^3+3 k \frac{l_{0}}{2 l_{1}^{5}}z^5+\mathcal{O}_{7}\biggl(\frac{z}{l_{1}}\biggr).
\end{equation}
and $f(z)\rightarrow -4 k z$ when $z\rightarrow \infty$. The oscillations generated by this nonlinear force on a mass $m$ with initial conditions $z(0)=0,z'(0)=-V_{0}, K_0= mV_{0}^{2}/2$  are described by the integral
\begin{equation}\label{eq:force4springs1}
t=C_{1} \pm \sqrt{\frac{m}{2}}\int_{0}^{z(t)} \frac{ds}{\sqrt{-2 k l_{1}^{2} -2k s^2 + 4k l_{0} \sqrt{s^2+l_{1}^{2}}}}.
\end{equation}
The RHS term in Eq. (\ref{eq:force4springs1}) is reducible to a sum of two elliptic integrals (on of the first kind and one incomplete) and the nonlinear oscillation can be expressed analytic in terms of special functions. In order to estimate in a first order of approximation the frequency of these oscillations we approximate the force with its cubic Taylor polynomial and we obtain a good approximation for the cubic nonlinear oscillations frequency of the payload inside the container
\begin{equation}\label{eq:frequency4springscubicapprox}
\nu [Hz]\sim \frac{2\sqrt{2k l_{1} (2 l_{0}-l_1)^{3/2}}}{z\biggl[2 (4 l_{0}-l_{1})+\frac{(l_{1}-l_{0})z^2}{3 l_{1}^{2}} \biggr]}.
\end{equation}

\section{Impact modeling}
\label{sec:impactmodeling}

In this section we write the dynamical equations for the system container plus payload under the action of the gravity, impact with ground modeled by the elastic force of the container Eq. (\ref{eq:forcesimpact1}), interaction between payload and container modeled by Eq. (\ref{eq:forceModel4Springs}), air resistance forces, and internal friction forces. We consider the motion 1-dimensional with 2 degrees of freedom. For a container of mass $M$ and a payload of mass $m$ we have the dynamical system
$$
M Z'' = -M g-F^{*}(L_{0}-Z)-f(z)-A Z^{'*} -\triangle^{*} \ \hbox{sign} (Z')-B Z^{'2*} \hbox{sign} (Z')
$$
\begin{equation}\label{eq:MainModelEq.Container}
-B_{a} Z^{'2} \hbox{sign} (Z'),
\end{equation}
\begin{equation}
(M+m) z'' = -m g+f(z)-\alpha z'-\delta \ \hbox{sign} \ (z'),
\label{eq:MainModelEq.Paylod}
\end{equation}
where $Z(t)$ is the height of the center of mass of the container with the origin taken on ground. By $L_0$ we denoted the distance between the center of mass of the container and the lowest contact point of the container, so this is the height at which the impact force starts to act upon the system. The parameters $A, B, \triangle$ are positive constants describing the drag coefficients for Stokes linear viscous force, Rayleigh quadratic viscous, and constant friction forces responsible of the container deformation. The star superscript shows that the quantity has that value only during the mechanic contact with the ground, and it is zero otherwise, if the container is airborne (that is $(x)^{*}=x $ if $0 \leq Z \leq L_{0}$ and is zero elsewhere). The parameter $B_{a}=\rho_{air} C_{D} \pi R_{\bot}^{2}/2$ describes the quadratic drag in air and is given by the geometry of the container.

A preliminary estimation of the air resistance of this container shows a terminal velocity in normal atmosphere of $3-8 m/s$ with parachute, and $24-30 m/s$ without. In both situations however, the Reynolds number ranges between $Re=2,500-150,000$ which exceeds the minimum limit of applying the Rayleigh formula for quadratic drag \cite{aerodyn}. However, during the impact with the ground the velocity of the container reduces and a Stokes linear term in velocity drag can become important during that phase. This is the reason for the specific choice of the damping terms in Eqs. (\ref{eq:MainModelEq.Paylod},\ref{eq:MainModelEq.Container}).
\begin{figure}[!b]
\centering
\centering\includegraphics[scale=0.9]{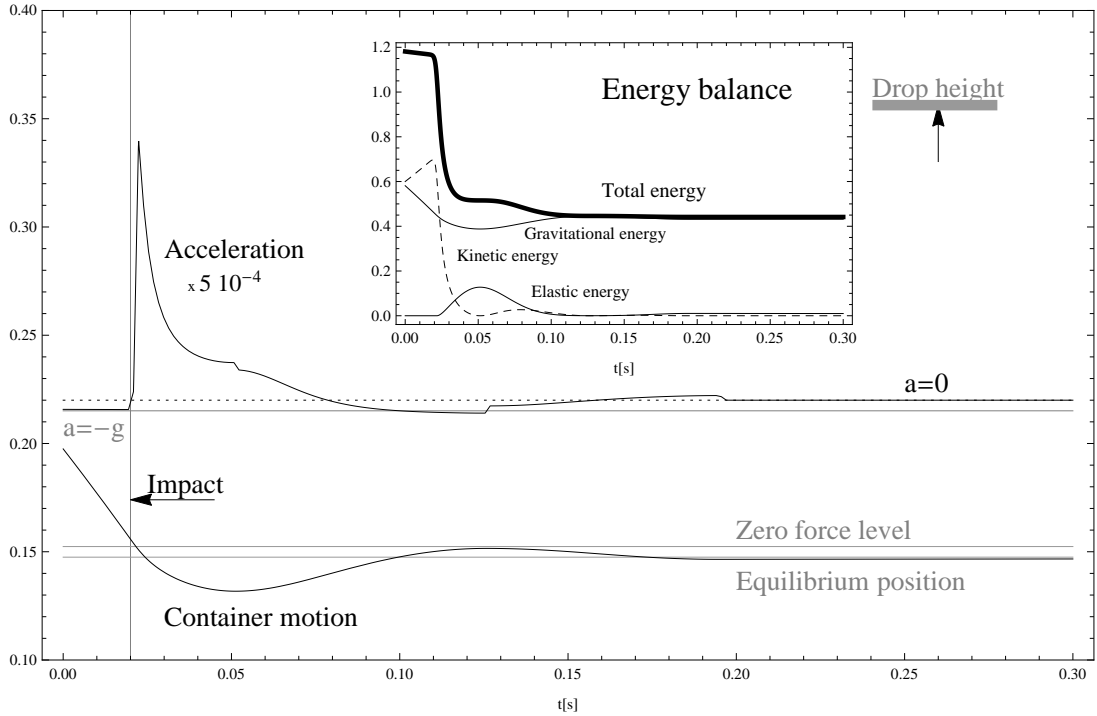}
\caption{Model calculations for impact of an empty container on cement. Lower curve shows motion of the sphere few milliseconds before the impact, and upper curve shows its acceleration (shifted upwards and re-scaled in this figure). In the inset we present energy balance between potential energy stored in the elastic force and gravitational and kinetic energy. The sum of all these energies is drawn with a thicker curve and shows two drops in energy: first steeper one by the visco-plastic impact deformation, and the second drop by visco-elastic deformation of the container's wall.}
\label{fig:justsphere}
\end{figure}

This nonlinear second order differential ordinary differential system  must be integrated for $t\geq 0$, where $0$ here denotes the impact moment, under initial conditions $Z(0)=L_{0}, z(0)=z_{eq}, Z'(0)=- V_{0} <0, z'(0)=0$. Here $z_{eq}$ is the equilibrium position of the payload inside the container at rest, that is the solution to the equation $f(z_{eq})=-mg$. This equation can be reduce to a quadric equation
$$
z^4+\frac{mg}{2k}z^3+\biggl( \frac{m^2 g^2}{16 k^2}+l_{1}^{2}-l_{0}^{2} \biggr) z^2+\frac{mg l_{1}^{2}}{2k}z+\frac{m^2 g^2 l_{1}^{2}}{16 k^2}=0.
$$
There is only one unique negative equilibrium solution $z_{eq}$ for this equation, and it can be approximated with
$$
z_{eq}\sim -\frac{mg}{4k}\biggl( 1+\sqrt{1-\frac{4 k^2 (l_{1}^{2}-l_{0}^{2})}{m^2 g^2}} \biggr)+\mathcal{O}_{2} ( mg/k ).
$$

\begin{figure}[!b]
\centering\includegraphics[scale=.7]{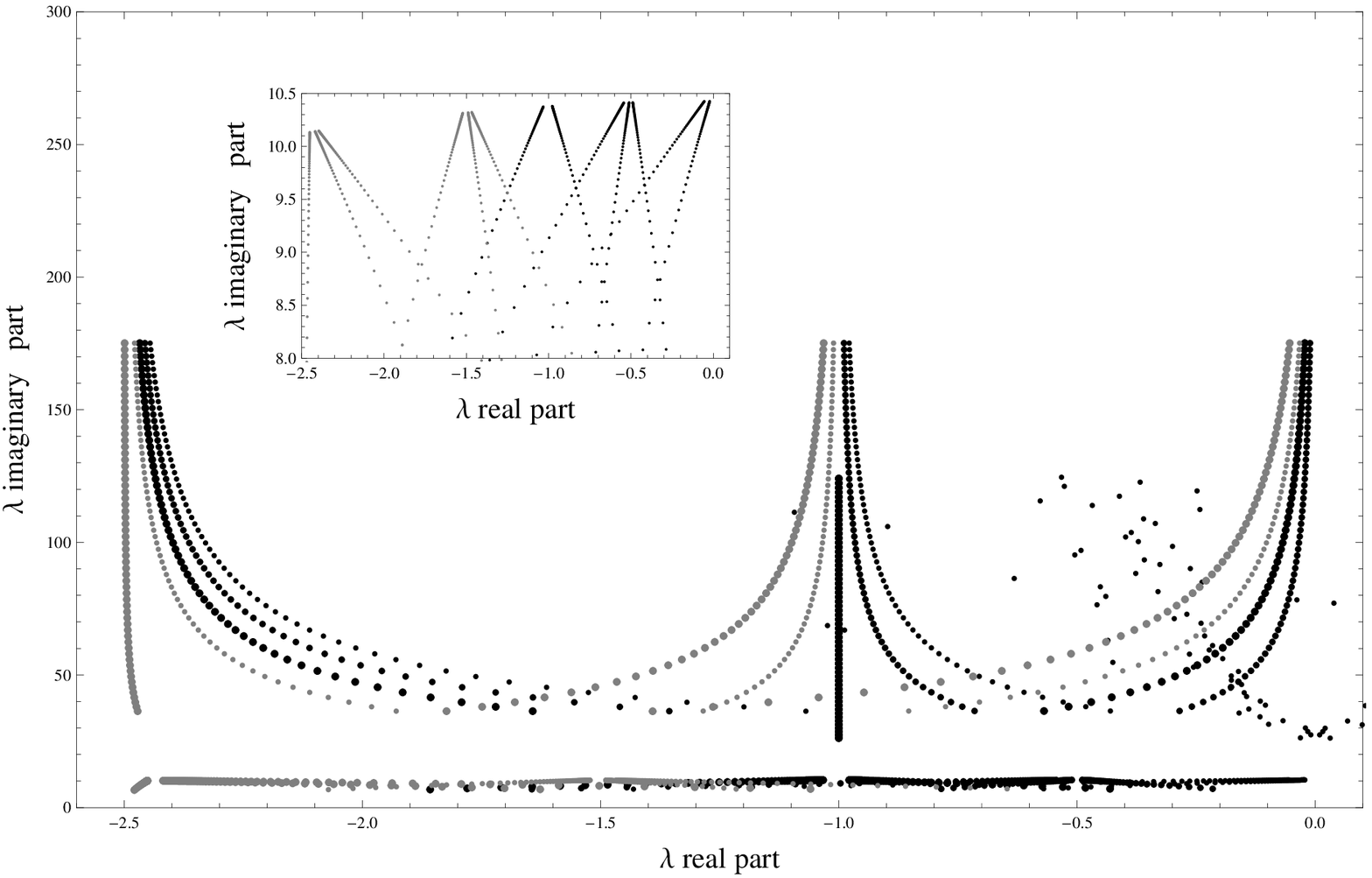}
\caption{Eigenvalues for the linearized system of differential equations for a set of parameters ranging $k=10^4$ N/m, $M=0.55$ Kg, $m=0.5$ Kg, $\alpha=0.0 \div 5 \cdot 10^{-8} \div 1.0 \div 5.0$ Ns/m, $A=0 \div5$ Ns/m, $\triangle=\delta=0$ N.}
\label{fig:diffgeomfluidA}
\end{figure}

\begin{figure}[!b]
\centering\includegraphics[scale=.7]{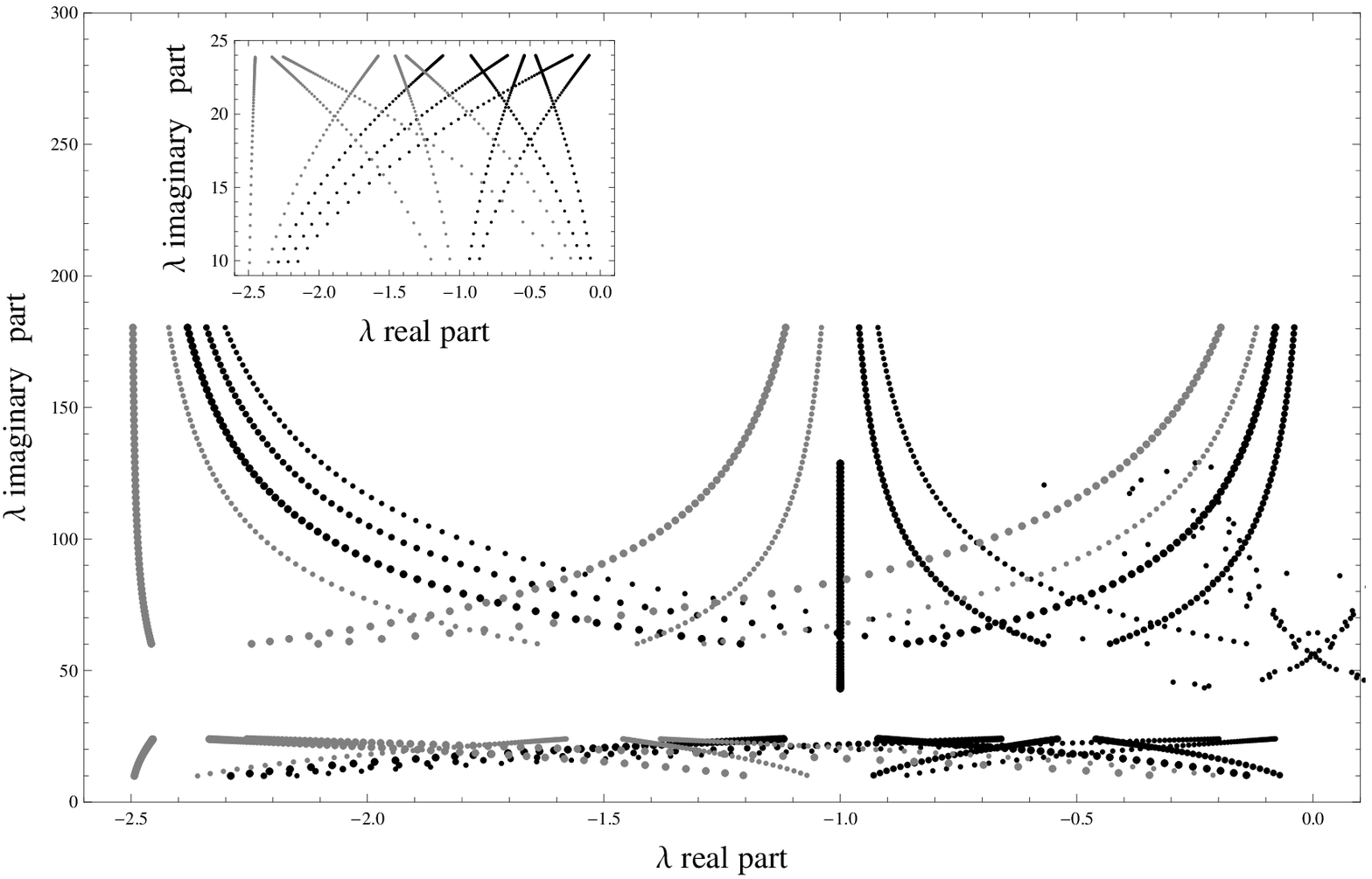}
\caption{Eigenvalues for the linearized system of differential equations for a set of parameters ranging  $k=5,000$ N/m, $M=2.0$ Kg, $m=0.5$ Kg, $\alpha=0-5 \cdot 10^{-8}-1-5$, $A=0-5$, $\triangle=\delta=0$ N. }
\label{fig:diffgeomfluidB}
\end{figure}

\begin{figure}[!b]
\centering\includegraphics[scale=.7]{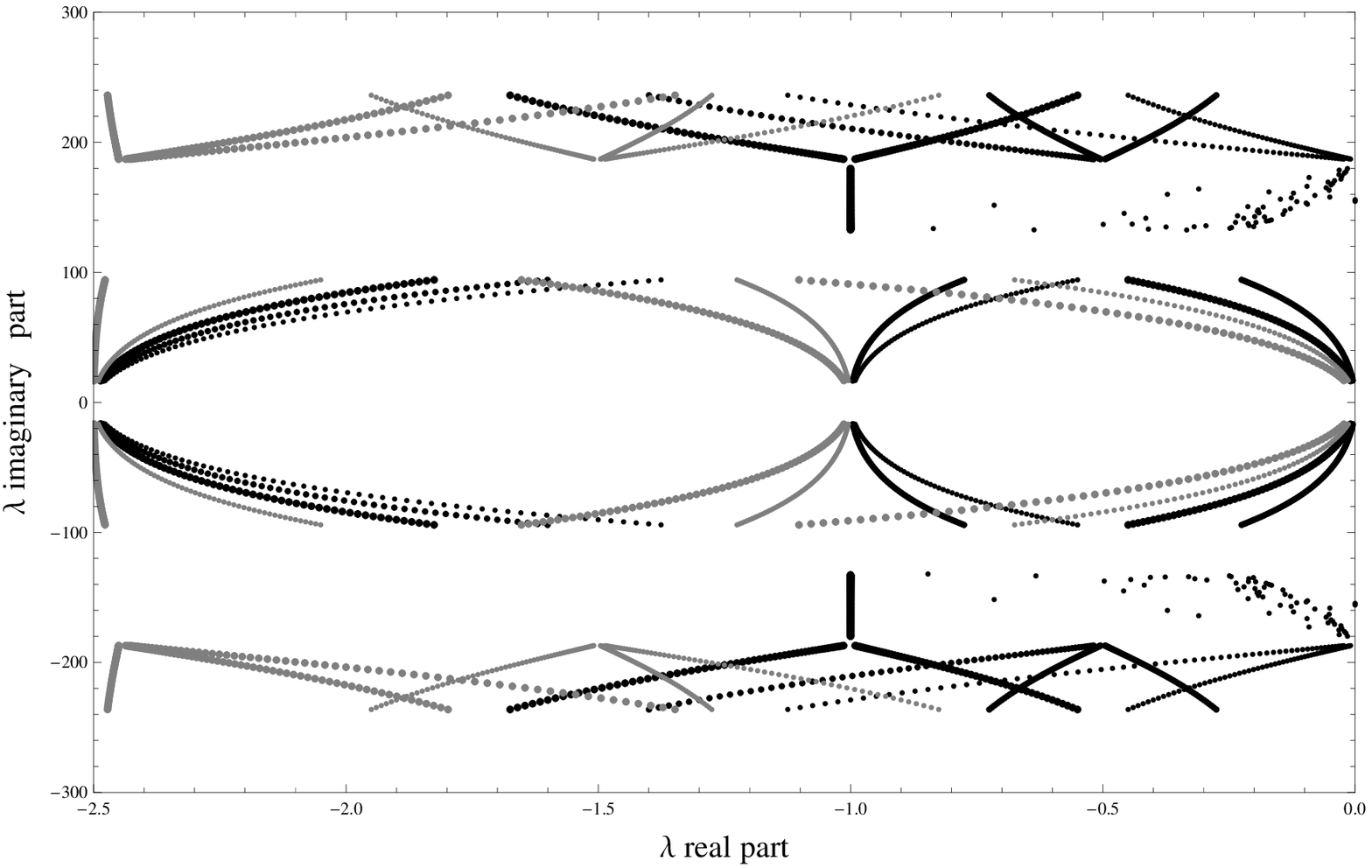}
\caption{Eigenvalues for the linearized system of differential equations for a set of parameters ranging  $k=10^4$ N/m, $M=0.55$ Kg, $m=0.5$ Kg, $\alpha=0-5 \cdot 10^{-8}-1-5$, $A=0-5$, $\triangle=\delta=0$ N.}
\label{fig:diffgeomfluidC}
\end{figure}

The linearization of the differential system around the equilibrium values
$$
Z''=z''0, \ \ Z'=z'=0, \ \ F(Z_{eq}-L_{0})+f(z_{eq})=-Mg, \ \ f(z_{eq}=-mg.
$$
conducts to an eigenvalue algebraic equation of the form
\begin{equation}\label{eq:CharactPolynomialEigevaluesLiniarized}
-\frac{k \epsilon}{M^2}+\biggl[ \lambda (\alpha+\lambda )+\frac{m}{M} \biggl(1+\frac{m}{M} \biggr) \frac{k}{m}  \biggr]\biggl(\frac{\epsilon}{M}+A\lambda +\lambda^2 \biggr)=0.
\end{equation}
The analysis of the eigenvalues is performed in Figs. \ref{fig:diffgeomfluidA}, \ref{fig:diffgeomfluidB}, and \ref{fig:diffgeomfluidC}.

\begin{figure}[!b]
\centering
\centering\includegraphics[scale=0.85]{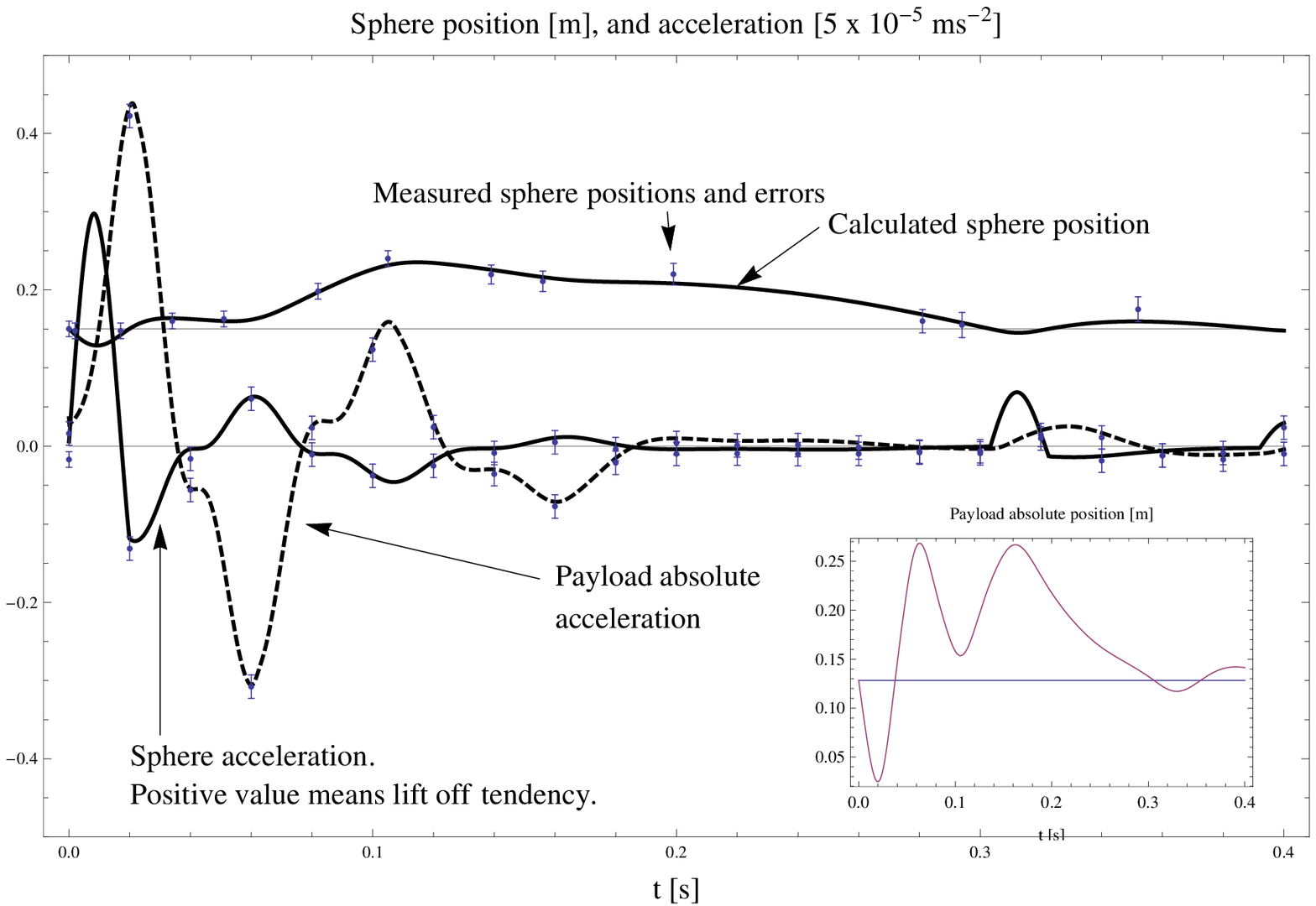}
\caption{Theoretical (solid curves)  and experimental (dots with error bars) results for a drop from $H=2.0 m$ on concrete, see also Fig.\ref{fig:rapidphotocexp3concr}. The upper solid line represents to motion of the center of mass of the container, beginning at the impact moment (labeled $t=0$) and placed at the container radius height $L_0$, reconstructed from the rapid camera images, Fig. \ref{fig:rapidphotocexp3concr}. The two lower curves represent the accelerations (solid for the container, and dashed for the payload) beginning at $t=0, g=-9.81 m/s^2$. In the inset we present the result of the integration of the payload acceleration data in order to obtain the motion of the payload with respect to the ground.}
\label{fig:theoryexp3concrete}
\end{figure}

In order to test the model we integrated numerically Eq. (\ref{eq:MainModelEq.Container}) for $m=0$. The result is presented in Fig. \ref{fig:theoryexp3concrete}.

\begin{figure}[!b]
\centering
\centering\includegraphics[scale=0.5]{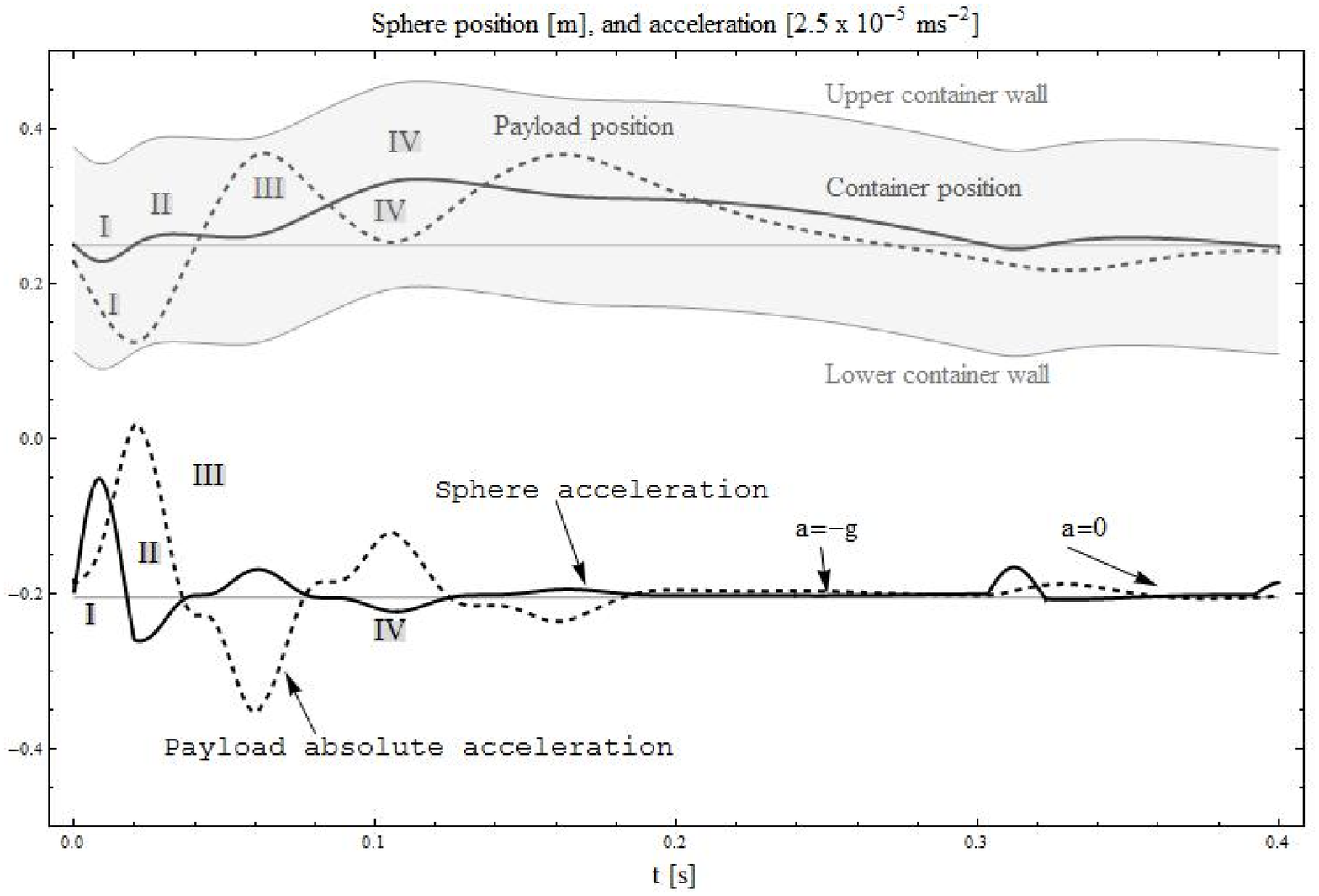}
\caption{Complete time line of the drop modeling presented in Fig. \ref{fig:theoryexp3concrete}. In the upper part of the frame we present the motion of the container center of mass (solid curve) and of the payload (dashed curve), both centered at heights $L_0$ and $L_0 - \delta L$, respectively. The gray stripe represents the motion of the upper and bottom wall of the container, and if the payload curve does not exit this stripe it means the payload did not collided into the container walls. In the lower part of the frame, we present the accelerations of the container (solid curve) and payload (dashed curve) both centered at their initial acceleration $-g$. The coupled motion can be explained as the repetition of four main phases labeled from $\mathbf{I}$ to $\mathbf{IV}$ and described in the text.}
\label{fig:theoryexp3concrete2}
\end{figure}
The motion of the container (sphere) $Z(t)$ a little before the impact, and after the impact and its acceleration are calculated. The energy balance shows a two-phase process of impact: in the first $15$ ms after impact the damping is dominated by the visco-plastic impact deformation, and the acceleration picks to its maximum value in the whole process. In the second stage, the process is longer ($ 60$ ms) and the damping is dominated by regular friction forces which burn the energy already stored in elastic force.

In Fig. \ref{fig:theoryexp3concrete} we present the theoretical calculations compared to the experimental results. In this case the container has $M=2.516 Kg$ out of which the payload was $m=0.507 Kg$, and it was dropped from $H=2.0m$ on concrete. The best model parameter fit were obtained for $k=740 N/m, K=61,340 N/m$, $\alpha=5.02 Kg/s, A=4.98 Kg/s, \delta =11.21 N$, and $\Delta=51.18 N$.

The relative errors for accelerations are smaller because of the precision of the electronic sensors, while the errors in the visual measurement of the container position slightly increase towards higher times because of the parallax error.

\begin{figure}[!b]
\centering\includegraphics[scale=1]{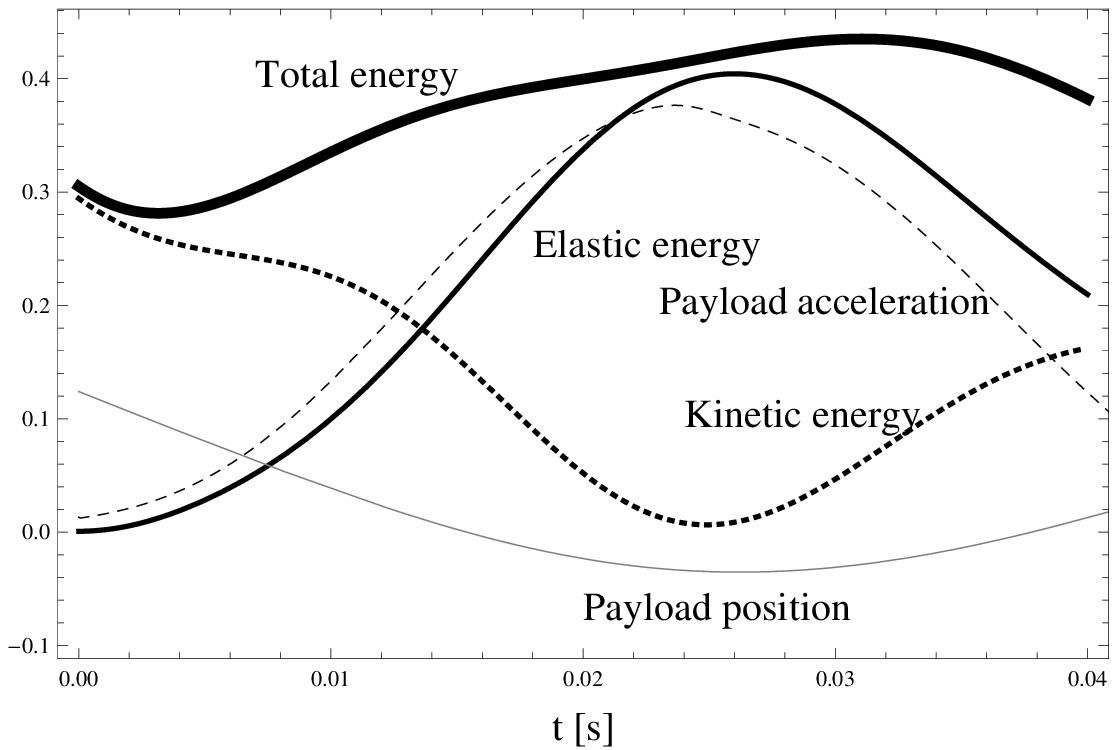}
\caption{Detail of the energy balance during bouncing off after impact.}
\label{fig:diffgeomfluid12}
\end{figure}

The experiment-theory match is good enough to induce confidence in the model, and an analysis of the stages of the impact process can be made. In the first stage (labeled $\mathbf{I}$ in Fig. \ref{fig:theoryexp3concrete2}) the container hits the ground, and its center of mass keeps descending a short interval of time because of the walls elasticity. In this stage the container experiences high acceleration with a maximum of about $600 g$. At the same time, the payload begins its trip downwards, also with high downwards acceleration, phase labeled $\mathbf{II}$, see also Fig. \ref{fig:rapidphotocexp3concr} (a). In this second phase, even if the container tends to bounce off the ground, its upward tendency of (see label $\mathbf{II}$ on top of the "Container position" solid curve) motion is slowed down by the springs opposite reaction. The acceleration of the payload fully develops, and the container is maintained at ground for a while.
After experiencing this large value, the payload acceleration changes the sign, and the payload moves upwards. The payload stops at a highest upper point, marked by $\mathbf{III}$ in Fig. \ref{fig:theoryexp3concrete2}, and also noticed in the frame (b) of Fig. \ref{fig:rapidphotocexp3concr}. Simultaneously, the container starts to feel the spring effect and is pulled upwards, too, see label $\mathbf{III}$ under sphere acceleration solid gray curve, Fig. \ref{fig:rapidphotocexp3concr} (c). In phase $\mathbf{IV}$ the container is lifted from ground at its maximum bouncing height, while the payload is in opposition of phase: its acceleration is oriented downwards, see frame (d) in Fig. \ref{fig:rapidphotocexp3concr}. The container bounces maximum twice times, while the payload performs 2.5 full oscillations. We notice that for the chosen configuration of the elastic coefficients and masses the damped container oscillations are somehow in opposition of phase with the payload, as one can see from the upper part in Fig. \ref{fig:theoryexp3concrete2} where the two curves representing their motion have always maxima in opposition.
\begin{figure}[!b]
\centering\includegraphics[scale=.9]{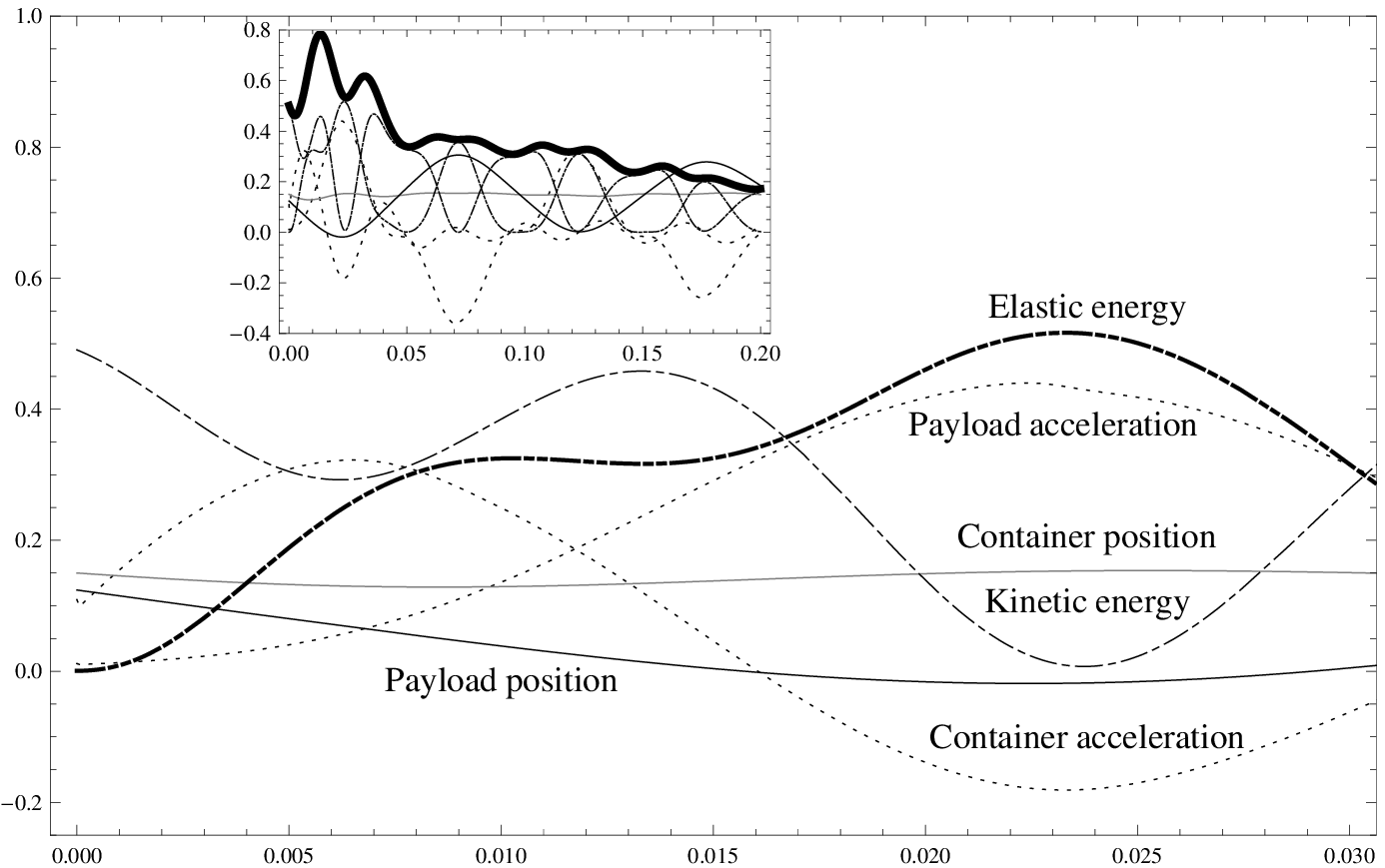}
\caption{Time evolution of the energies involved in the impact in comparison with motion and acceleration.  In the inset we present a longer time scale which shows the rate of dissipation of energy (thick solid curve).}
\label{fig:diffgeomfluid1q}
\end{figure}
In Fig. \ref{fig:diffgeomfluid1q} we present the energy balance in one typical impact. An interesting observation must be mentioned because the phenomenon repeats almost in all situations. The first acceleration stage right after impact the container and payload accelerations are in opposition of phase, and they attain their maximum accelerations at minimum total kinetic energy, and maximum total potential energy, see Fig. \ref{fig:diffgeomfluid12}, too.
The study in Fig. \ref{fig:diffgeomfluid12} describes a container of radius $R=0.127$ and mass $M=1$ Kg, with a payload  $m=0.5$ Kg, attached with springs of characteristics $l_{1}=l_{0}=0.09$ m, $k=1200$ N/m, $\epsilon=6000$ N/$m^{1/2}$ and $L_{0}=0.15$ m. The best fit indicates damping parameters $\alpha=0.5$ Ns/m, $\delta=8$ N, $A=80.2$ Ns/m, and $\triangle=1.13$ N.

\begin{figure}[!b]
\centering\includegraphics[scale=.7]{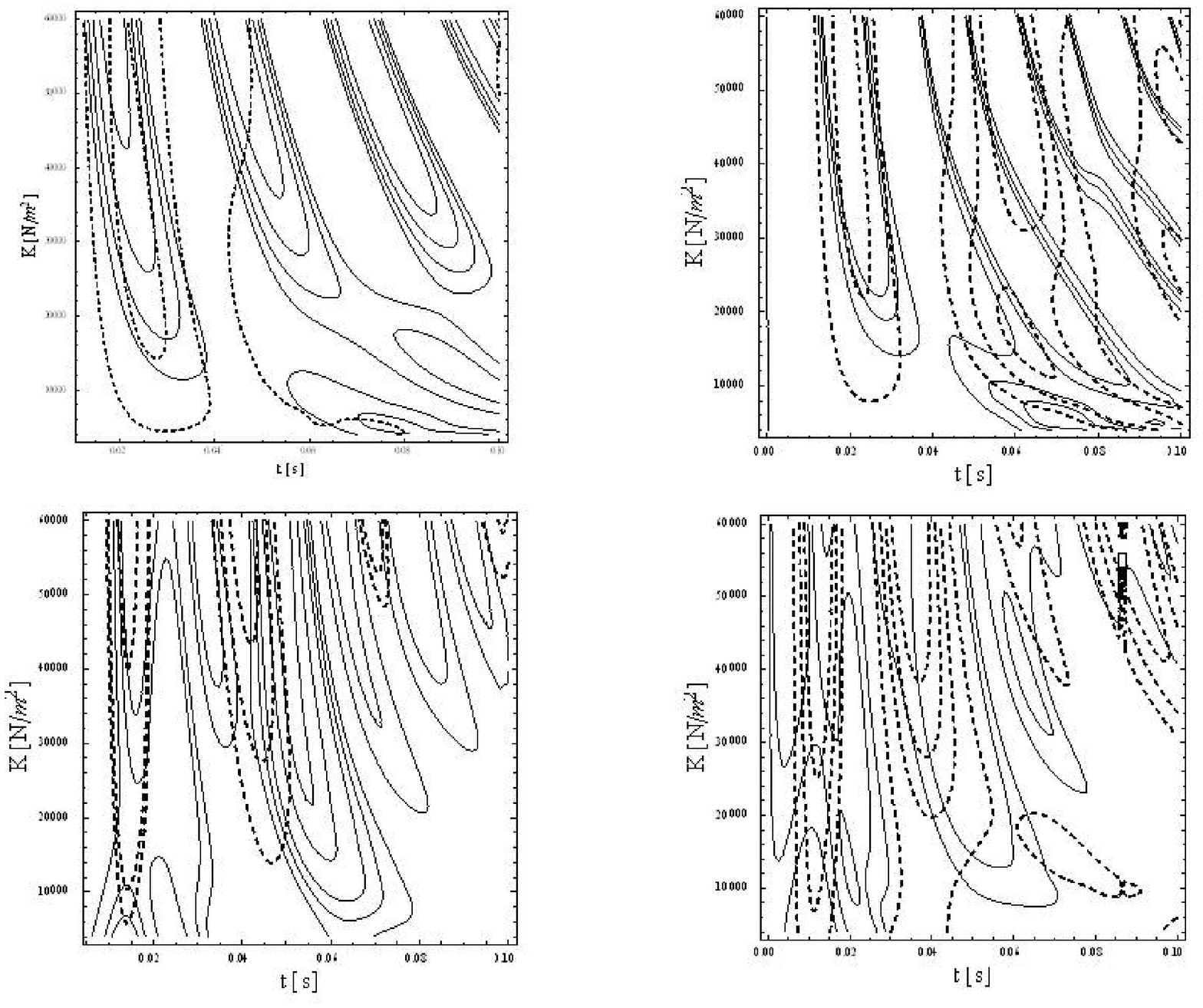}
\caption{Contour plots of the acceleration of the container ($Z''$, dashed curves) and of the payload ($z''+Z''$, solid curves) function of time, for different values of the elastic constant $k$. In all the cases we choose $M=1$ Kg, $m=0.5$ Kg, $l_{0}=l_{1}=0.09$ m, $R=1.4 l_{0}$, and $L_0 =0.17$ m. The differences between the figures are, clockwise from top left: \textit{Upper left=} $\epsilon=3500$  N/$m^{1/2}$, $H=1$ m and no dissipation or drag.  \textit{Upper right=}  $\epsilon=3500$  N/$m^{1/2}$, $H=10$ m, and no dissipation or drag. \textit{Bottom left=} $\epsilon=6000$  N/$m^{1/2}$, $H=1$ m, $\alpha=0.5$ Ns/m, $A=80$ Ns/m, $\delta=8.0$ N, and $\triangle=400$ N. \textit{Bottom right=} $\epsilon=6000$  N/$m^{1/2}$, $H=10$ m, $\alpha=0.5$ Ns/m, $A=80$ Ns/m, $\delta=8.0$ N, and $\triangle=400$ N.}
\label{fig:diffgeomfluid342534}
\end{figure}

The most interesting result we noticed is that the nonlinear oscillations of the payload inside the container, and the container motion are almost always in opposition of phase, that is one is maximum when the other is minimum. This effect seems to be independent of the initial conditions (drop height) the parameters of the container or payload, or the dissipation coefficients. We present such a study in Fig. \ref{fig:diffgeomfluid342534} where we plotted contour plots of the acceleration of the container and the absolute acceleration of the payload on the same  $(t,k)$ plane. The horizontal direction in these four plots is time line, and the vertical direction shows different values for the elastic constant of the payload connecting springs, $k$.
The two upper windows in Fig. \ref{fig:diffgeomfluid342534} represent systems without dissipation, left being dropped from low height, and right frame from high height. The bottom row of frames represent the same height (left and right) as above, except the system has regular dissipation, drag and friction as usual. WE note that the center lines of the contour levels always coincides in the first oscillation. This is a strong argument in favor of the synchronization of the two motions. In other words, we have a lock in of the phase of the two sub-systems, which guarantees in any situation (any drop height, any terminal velocity, any initial energy, any dissipation parameters, etc.) the minimal acceleration of the payload, hence its protection.

\section{Study of the influence of the ground type}
\label{sec:last}

\begin{figure}[!b]
\centering\includegraphics[scale=0.5]{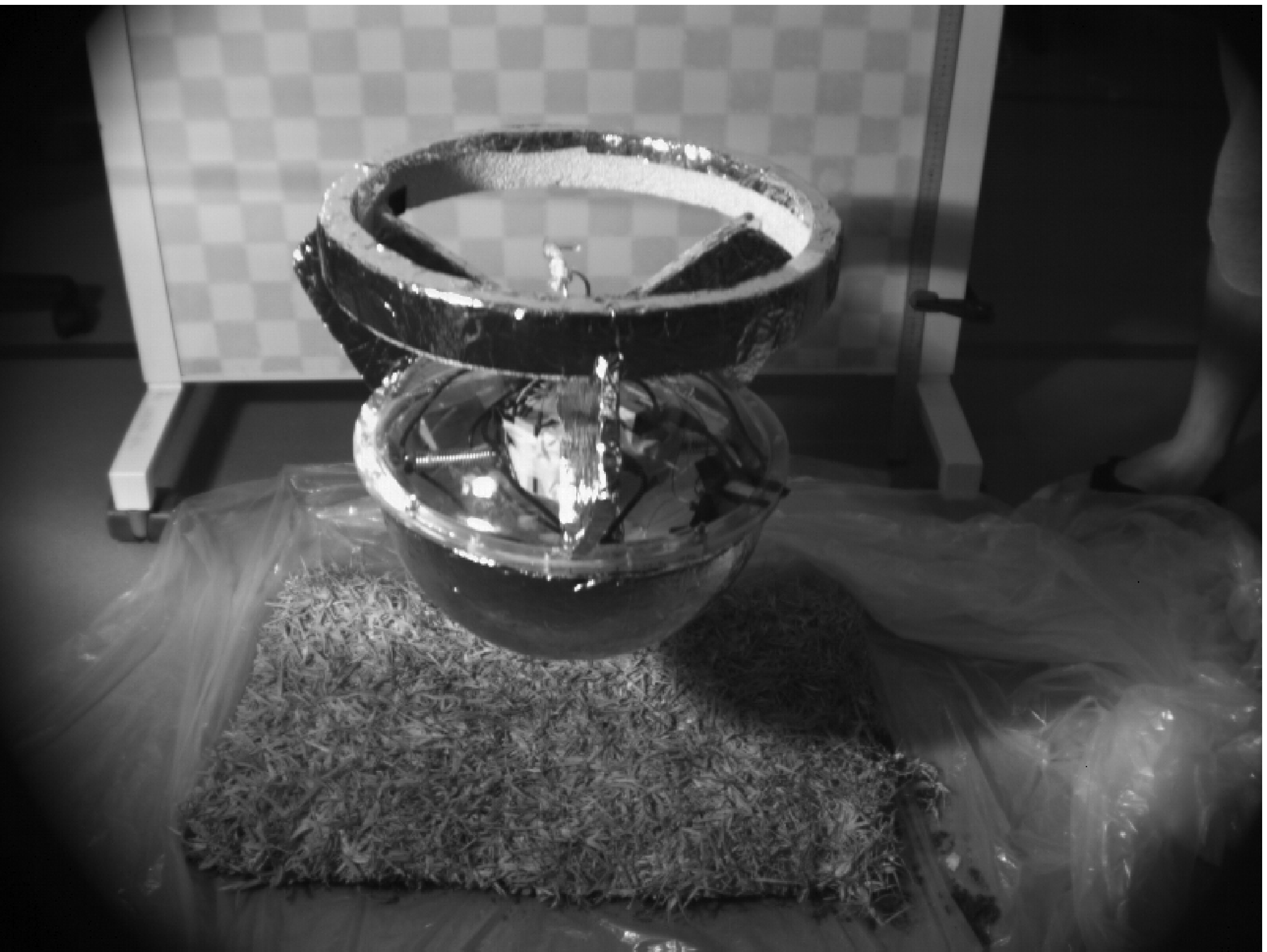}
\caption{Impact on grass.}
\label{fig:diffgeomfluid5879464573}
\end{figure}

\begin{figure}[!b]
\centering\includegraphics[scale=0.5]{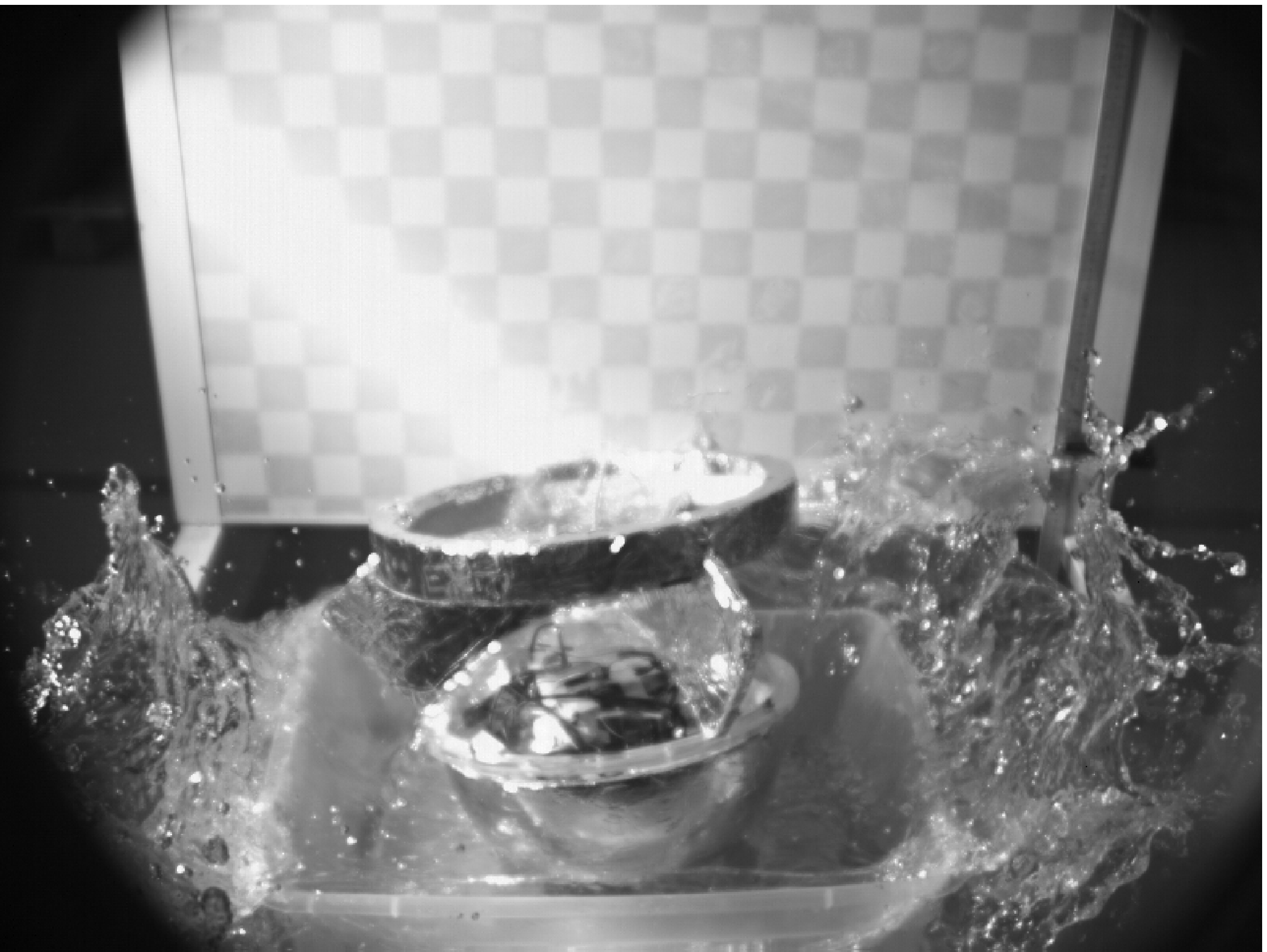}
\caption{Waves and surge created by impact on water.}
\label{fig:diffgeomfluidwfewgvgt}
\end{figure}

We experimented four types of drops on different grounds: concrete, sand, grass (for example Fig. \ref{fig:diffgeomfluid5879464573}) and water (for example Fig. \ref{fig:diffgeomfluidwfewgvgt})

\begin{figure}[!b]
\centering\includegraphics[scale=0.8]{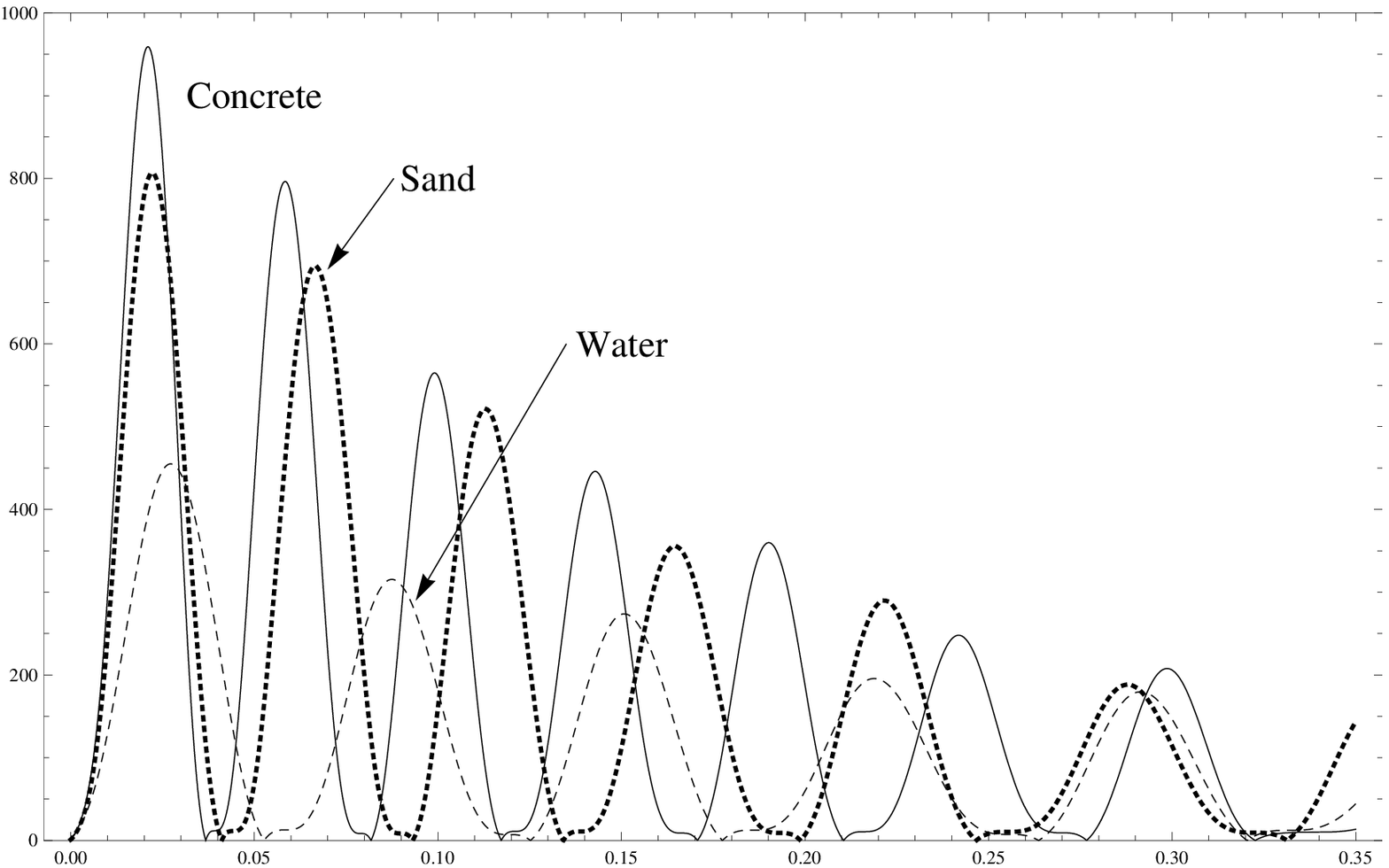}
\caption{The energy dissipation profiles versus time for different types of ground.}
\label{fig:didmjfdjruifjhd}
\end{figure}

There are differences in the behavior of the accelerations, but also similarities. In Fig. \ref{fig:didmjfdjruifjhd} we present the time profiles of the energy for three types of ground. One can note a different rate of releasing the energy as well as phase shift in the oscillations. At the same time, the inherent nature of the process remain independent of the type of ground.

One can notice the same degree of synchronism of the maxima and minima, both in position and acceleration for all types of impacts. We calculated the overlap between the velocities profiles of the container and payload
\begin{equation}\label{eq:444ut}
\int_{t=0}^{T_{payload}}Z'(t) \cdot z'(t) dt,
\end{equation}
on a period of the payload oscillation, which expression is actually a measurement of the degree of asynchronism of the relative motion of the payload and the container motion. The results are presented in the Table \ref{tab:title1}.
\begin {table}[H]
\caption {The degree of synchronism between the two oscillations, Eq.(\ref{eq:444ut}), container and payload, for different impact velocities.} \label{tab:title1}
\begin{center}
\begin{tabular}{|c|c|c|c|c|}
\hline
Ground type: & Concrete & Grass & Sand & Water \\
\hline
$V_0 =2 $ m/s & 0.848 & 47.7 & 1.36 & 24.2 \\
\hline
$V_0 =10$ m/s & -14.3& 49.3 & 2.21 & 29.9 \\
\hline
$V_0 =15$ m/s & -26.2 & 50.4 & 4.04 & 30.8 \\
\hline
\end{tabular}
\end{center}
\end{table}

In Table \ref{tab:title2} we present the  maximum values of the payload absolute acceleration measured in g's for different types of ground. The values are taken from experiments, and verified for match with the numerical code results and  errors less then $10\%$.
\begin {table}[H]
\caption {Maximum absolute acceleration $\max |Z''(t)+z''(t)|$ of the payload presented in units of $g$.} \label{tab:title2}
\begin{center}
\begin{tabular}{|c|c|c|c|c|}
\hline
Ground type: & Concrete & Grass & Sand & Water \\
\hline
$V_0 =2 $ m/s & 24.9 & 18.3 & 19.3 & 8.39 \\
\hline
$V_0 =10$ m/s & 105 & 77.0 & 54.2 & 61.3 \\
\hline
$V_0 =15$ m/s & 257 & 159 & 126 & 142 \\
\hline
\end{tabular}
\end{center}
\end{table}

\section{Conclusions}

In this study we investigated the dynamics of the impact of a solid container and a payload by dropping from different heights (from $H=0.1$ m to $H=10$ m) on different types of ground (concrete, sand, grass and water). We ran experiments and measure instantaneous acceleration of both the container and the payload inside, and we taped the motion of the container using a rapid photography camera and take the image against a  chessboard wall. We also elaborated a nonlinear one dimensional two-degrees of freedom model to simulate the evolution and calculate accelerations and energies. The comparison between the experiment and theory was very good. For the special system of nonlinear springs we designed we noticed a similar type of behavior of the payload for all types of ground in which the amplitude and acceleration of the payload is always in opposition of phase with the ones of the container. This phenomenon reveals the possibility of the existence of a energy resonant damping which allows a faster transfer of the shock energy from the payload back to the container. This may be the signature of a non-linear energy sink process and this topic will be studied in a forthcoming paper.

\end{document}